# Superconductivity in Co-doped SmFeAsO

Yanpeng Qi, Zhaoshun Gao, Lei Wang, Dongliang Wang, Xianping Zhang, Yanwei Ma[*]

Key Laboratory of Applied Superconductivity, Institute of Electrical Engineering,

Chinese Academy of Sciences, P. O. Box 2703, Beijing 100190, China

**Abstract:**

Here we report the synthesis and characterizations of $SmFe_{1-x}Co_xAsO$ (x=0.10, 0.15) for the first time. The parent compound SmFeAsO itself is not superconducting but shows an antiferromagnetic order near 150 K, which must be suppressed by doping before superconductivity emerges. With Co-doping in the FeAs planes, antiferromagnetic order is destroyed and superconductivity occurs at 15.2 K. Similar to $LaFe_{1-x}Co_xAsO$, the $SmFe_{1-x}Co_xAsO$ system appears to tolerate considerable disorder in the FeAs planes. This result is important, suggesting different mechanism between cuprate superconductors and the iron-based arsenide ones.

[*] Author to whom correspondence should be addressed; E-mail: ywma@mail.iee.ac.cn

Recent discovery of high-$T_c$ superconductivity in iron pnictides has generated highly intensive research activities in solid state physics [1-9]. After first report on LaFeAsO$_{1-x}$F$_x$ with critical temperatures $T_c$ of 26 K [1], even higher transition temperature up to 55 K in SmFeAsO$_{1-x}$F$_x$ followed quickly，which is the first non-copper-oxide superconductor with $T_c$ exceeding 50 K. The system attracts much attention not only from high transition temperature but also from a point that many related systems can be derived by the substitution of the constituent elements. The parent compound ReFeAsO (Re=rare earth element) itself is not superconducting but shows an anomaly at around 150 K, electron doping by F suppresses the anomaly and recovers the superconductivity. Theoretical calculations indicated the itinerant character of Fe 3d electrons in the iron-based superconductors [10, 11]. In order to get deeper insight into 3d electron and find the original of superconductivity, it is necessary to make a doping experiment on the Fe site, since doping was generally performed on sites in-between the Fe-As layers presently, either on the Re-site or on the O-site in the ReFeAsO compounds. Recently, Sefat et al. reported the superconductivity in Co-doping LaFeAsO [12] and BaFe$_2$As$_2$ [13]. Leithe-Jasper et al. reported the superconductivity with $T_c$ up to 20 K in the SrFe$_{2-x}$Co$_x$As$_2$ [14]. Despite the new insights provided by these studies, the question of high-temperature superconductivity continues to elude description by physical models. We consider that the detailed investigation of Co-doping in the FeAs planes would give more important clues to understand the mechanisms on iron-based superconductor. Here we report that superconductivity was realized by doping magnetic element cobalt into the superconducting-active FeAs layers in SmFeAsO. The antiferromagnetic spin-density-wave transition in the parent compound was suppressed, and superconductivity with $T_c \sim 15$ K was induced. This result is important, indicating essential difference between cuprate superconductors and the iron-based arsenide ones.

The synthesis of Co-doped SmFeAsO was carried out by one-step solid state reaction. The details of fabrication process are described elsewhere [9]. Stoichiometric amounts of the starting elements Sm, Co$_2$O$_3$, Fe, Fe$_2$O$_3$ and As were thoroughly grounded by hand and encased into pure Nb tubes. After packing, this tube was subsequently rotary swaged and sealed in a Fe tube. The sealed samples were heated at 1180 $^o$C for 45 hours. The high purity argon gas was allowed to flow into the furnace during the heat-treatment process. The sintered samples were obtained by breaking the Nb tube. It is noted that the sample preparation process except for annealing was performed in a glove box under high pure argon.

Phase identification and crystal structure investigation were carried out using x-ray diffraction (XRD) using Cu K$_\alpha$ radiation. Resistivity measurements were performed by the conventional four-point-probe method. AC magnetic susceptibility of the samples was measured by a Quantum Design physical property measurement system (PPMS).

Figure 1 shows XRD patterns for the prepared samples. It is seen that all main peaks can be well indexed based on the ZrCuSiAs tetragonal structure, indicting that the samples are single phase. The lattice parameters are $a$=3.9412 Å, $c$ = 8.4802 Å for the sample with x = 0.10, while $a$ = 3.9411 Å, $c$ = 8.4638 Å for the sample with x = 0.15. Clearly Co-doping leads to an obvious decrease in $c$-axis lattice while the $a$-axis remains nearly unchanged. Similar behavior is observed in the LaFe$_{1-x}$Co$_x$AsO [15]. Compared to the parent compound SmFeAsO, the apparent reduction of the lattice volume upon Co-doping indicates a successful chemical substitution. Small amount of impurity phases, mostly perhaps FeAsO, were also observed in the XRD patterns. Such impurity phases might be reduced by optimizing the heating process and stoichiometry ratio of start materials.

Figure 2 shows the temperature dependence of resistivity for SmFe$_{1-x}$Co$_x$AsO samples. As reported by Chen et al., undoped SmFeAsO sample exhibits a clear anomaly near 150 K [2], which is ascribed to the spin-density-wave instability and structural phase transitions from tetragonal to orthorhombic symmetry. For SmFe$_{0.9}$Co$_{0.1}$AsO, the room temperature resistivity $\rho_{300 K}$ =2.6 mΩ cm, this value is much smaller than the parent sample. As seen from Fig.2, the resistivity of the x = 0.1 sample decreases slowly with decreasing temperature, then the resistivity increases below 120 K, which is similar to undoped sample, and finally we can observe a rapid transition with the onset temperature 15.2 K, indicating a good quality of our samples. Up to 15% doping, the overall resistivity decreases obviously, while the transition temperature keeps almost unchanged with different doping levels, implying the essential of this superconductor. Compared to $SmFeAsO_{1-x}F_x$, the transition temperature is significantly lower, which is likely due to the stronger effect of disorder induced by doping in the FeAs layers. It is noted that the Co content for superconductivity is even smaller than that of F content, demonstrating that Co-doping is effective and strongly suppresses the antiferromagnetic order.

In order to further confirm the superconductivity of $SmFe_{1-x}Co_xAsO$, AC susceptibility measurement was also performed. Figure 3 shows the temperature dependence of AC magnetization for the $SmFe_{0.9}Co_{0.1}AsO$ sample. The sample shows a good diamagnetic signal. The onset critical temperature is 14.2 K, which is well corresponding to the middle transition point of resistance. The sharp magnetic transitions of AC curves suggest the good quality of our superconducting samples.

It is very interesting that superconductivity was realized by doping magnetic element cobalt into the superconducting-active FeAs layers, which would give more important clues to understand the mechanisms on iron-based superconductor. Firstly, it is surprising that superconductivity is induced by Co-doping, which challenges our previous understanding on superconductivity theory. It is known that there are a lot of examples in which superconductivity occurred by chemical substitution, such as Ba-doped $La_2CuO_4$ [16], K-doped $BaBiO_3$ [17], F-doped ReFeAsO [1], K-doped $BaFe_2As_2$ [18], etc. It should be noted that all the above dopants are non-magnetic, since superconductivity is not compatible with magnetism and magnetic atoms generally break superconducting Cooper pairs. However, cobalt is a typical magnetic element and the evidence of Co-doping induced superconductivity is in contrary to all previous assumptions about the competition between superconductivity and magnetic moments. Therefore, there should be some underlying mechanisms we can not understand completely at present. Secondly, the Fe-As layer is usually thought to be responsible for superconductivity and Re-O layer is carrier reservoir layer to provide electron carrier in ReFeAsO compounds. Similar to the cuprate superconductors, superconductivity can also be induced by charge doping from a reservoir layer in ReFeAsO compounds. In addition, Fe and Cu are both 3d element, so an analogy between the high temperature superconductor in the cuprates and the iron arsenide layer compounds was suggested by large number of reports [12-15]. However, recently superconductivity was induced by Co-doping in LaFeAsO [12] and SrFe$_2$As$_2$ [14] and here we report superconductivity occurred in SmFeAsO compound by substitution of Co for Fe. The relatively high $T_c$ occurrence by doping on the FeAs conducting layers clearly demonstrates that the in-plane disorder is highly tolerated in SmFeAsO compound. This result is different from cuprate superconductors, in which superconductivity is always damaged by doping on CuO$_2$ planes, thus further demonstrating that an analog with the high temperature superconductor in the cuprates is not appropriate. Therefore the model and mechanism of two class high temperature superconductors should be different. To investigate the origin of this behavior, further experimental and theoretical studies are required.

To summarize, we have successfully synthesized the iron-based Co-doped layered compound SmFe$_{1-x}$Co$_x$AsO by one-step solid state reaction method. Co-doping is effective and superconductivity is observed at 15 K**.** Similar to LaFe$_{1-x}$Co$_x$AsO, the SmFeAsO system appears to tolerate considerable disorder in the FeAs planes, suggesting the difference between cuprates and the iron-based arsenide ones. Our data demonstrates that an analogy between the high temperature superconductor in the cuprates and in the iron arsenide layer compounds is not really appropriate. Co-doping in the FeAs planes is interesting, which would give us important clues to understand the mechanisms on iron-based superconductor.

**Acknowledgement**

The authors thank Profs. Haihu Wen, Liye Xiao and Liangzhen Lin for their help and useful discussions. This work is partially supported by the Beijing Municipal Science and Technology Commission under Grant No. Z07000300700703, National '973' Program (Grant No. 2006CB601004) and National '863' Project (Grant No. 2006AA03Z203).

# References

[1] Kamihara Y., Watanabe T., Hirano M. and Hosono H, J. Am. Chem. Soc. **130**, 3296 (2008).

[2] Chen X. H., Wu T., Wu G., Liu R. H., Chen H. and Fang D. F., Nature **453**, 376 (2008).

[3] Wen H. H., Mu G., Fang L., Yang H., and Zhu X., Europhys. Lett. **82**, 17009 (2008).

[4] Chen G. F., Li Z., Wu D., Li G., Hu W. Z., Dong J., Zheng P., Luo J. L. and Wang N. L., Cond-mat: arXiv, 0803.3790 (2008).

[5] Ren Z. A., Yang J., Lu W., Yi W., Che G. C., Dong X. L., Sun L. L. and Zhao Z. X., Cond-mat: arXiv, 0803.4283 (2008).

[6] Ren Z. A., Yang J., Lu W., Yi W., Shen X. L., Li Z. C., Che G. C., Dong X. L., Sun L. L., Zhou F. and Zhao Z. X., Europhys. Lett. **82**, 57002 (2008).

[7] Ren Z. A., Yang J., Lu W., Yi W., Shen X. L., Li Z. C., Che G. C., Dong X. L., Sun L. L., Zhou F. and Zhao Z. X., Chin. Phys.Lett. **25**, 2215 (2008).

[8] Cheng P., Fang L., Yang H., Zhu X. Y., Mu G., Luo H. Q, Wang Z. S. and Wen H. H., Science in China G **51**, 719 (2008).

[9] Ma Y. W., Gao Z. S., Wang L., Qi Y. P., Wang D. L. and Zhang X. P., Cond-mat: arXiv, 0806.2839 (2008).

[10] Singh D.J. and Du M.H., Phys. Rev. Lett. **100**, 237003 (2008).

[11] Haule K., Shim J.H. and Kotliar G., Phys. Rev. Lett. **100**, 226402 (2008).

[12] Sefat A. S., Huq A., McGuire M. A., Jin R. Y., Sales B. C., Mandrus. D., Cranswick L. M. D., Stephens P. W. and Stone K. H., Cond-mat: arXiv, 0807.0823 (2008).

[13] Sefat A. S., McGuire M. A., Jin R. Y., Sales B. C. and Mandrus. D., Cond-mat: arXiv, 0807.2237 (2008).

[14] Leithe-Jasper A., Schnelle W., Geibel C. and Rosner H., Cond-mat: arXiv, 0807. 2223 (2008).

[15] Wang C., Zhu Z. W., Jiang S., Chi S., Luo Y. K., Ren Z., Tao Q., Wang Y. T., Cao

G. H. and Xu Z. A., Cond-mat: arXiv, 0807. 1304 (2008).

[16] Bednorz J. G. and Muller K. A., Z. Phys. B. **64**,189 (1986).

[17] Cava R. J., Batlogg B., Krajewski J. J., Farrow R., Rupp L. W., White A. E., Short K., Peck W. F. and Kometani T., Nature **332**, 814 (1988)

[18] Rotter M. Tegel M. and Johrendt D., Cond-mat: arXiv, 0805. 4630 (2008).

**Captions**

Figure 1 XRD patterns of the $SmFe_{1-x}Co_xAsO$ samples. The impurity phases are marked by *.

Figure 2 Temperature dependence of resistivity for the $SmFe_{1-x}Co_xAsO$ samples measured in zero field. Inset: Enlarged view of low temperature, showing superconducting transition.

Figure 3 Temperature dependence of magnetic susceptibility for the $SmFe_{0.9}Co_{0.1}AsO$ sample measured with $H_{ac} = 0.1$ Oe, $f = 333$ Hz.

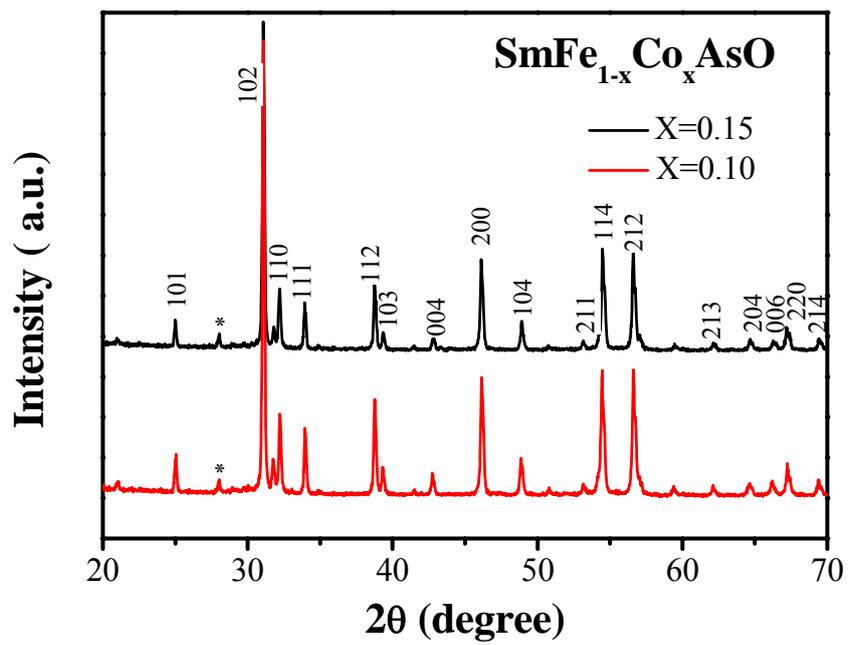

Fig.1 Qi et al.

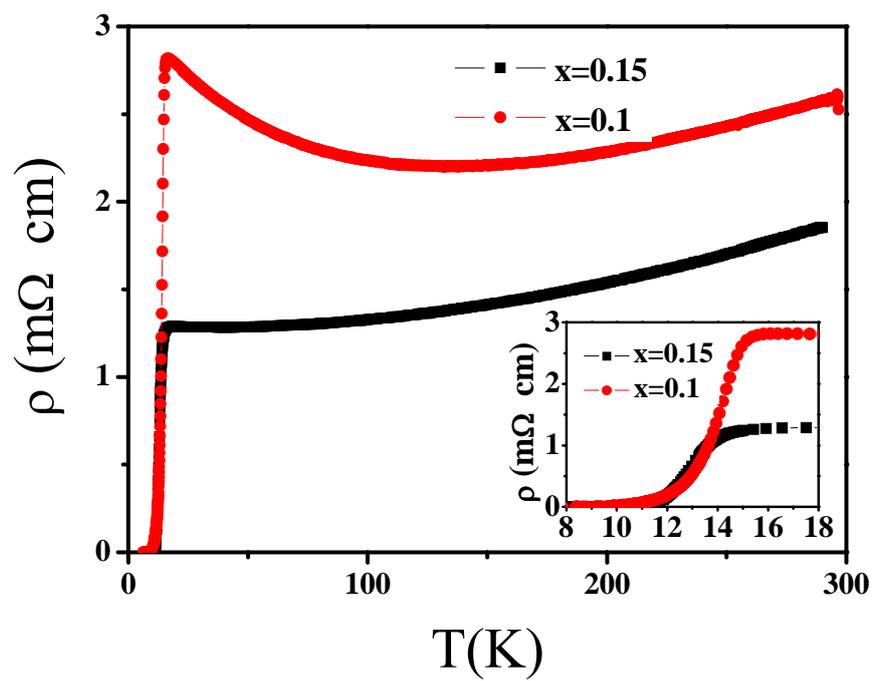

Fig.2 Qi et al.

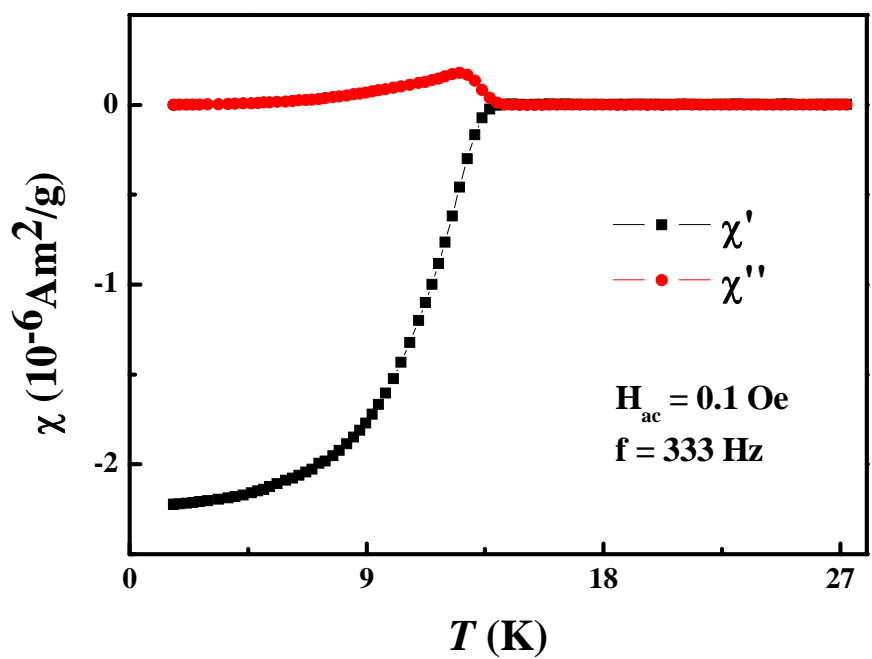

Fig.3 Qi et al.